\documentclass[runningheads]{llncs}
\usepackage{graphicx}

\usepackage[normalem]{ulem}
\useunder{\uline}{\ul}{}
\usepackage{adjustbox}
\usepackage{multirow}

\begin{document}
\title{Inferring Human Traits From Facebook Statuses}

\author{Andrew Cutler \and
Brian Kulis}
%
%
\institute{Boston University, Boston, MA}

\maketitle              
\vspace{-10pt}
\begin{abstract}
This paper explores the use of language models to predict 20 human traits from users' Facebook status updates. The data was collected by the myPersonality project, and includes user statuses along with their personality, gender, political identification, religion, race, satisfaction with life, IQ, self-disclosure, fair-mindedness, and belief in astrology. A single interpretable model meets state of the art results for well-studied tasks such as predicting gender and personality; and sets the standard on other traits such as IQ, sensational interests, political identity, and satisfaction with life. Additionally, highly weighted words are published for each trait. These lists are valuable for creating hypotheses about human behavior, as well as for understanding what information a model is extracting. Using performance and extracted features we analyze models built on social media. The real world problems we explore include gendered classification bias and Cambridge Analytica's use of psychographic models.

\end{abstract}

\section{Introduction}
Facebook's 2 billion users spend an average of 50 minutes a day on Facebook, Messenger, or Instagram \cite{stewart2016facebook}. Industry seeks to obtain, model and actualize this mountain of data in a variety of ways. For example, social media can be used to establish creditworthiness \cite{suncorp,khandani2010consumer}, persuade voters \cite{cogburn2011networked,gonzalez2017hacking}, or seek cognitive behavioral therapy from a chatbot \cite{fitzpatrick2017delivering}. Many of these tasks depend on knowing something about the personal life of the user. When determining the risk of default, a creditor may be interested in a debtor's impulsiveness or strength of support network. A user's home town could disambiguate a search term. Or---reflecting society's values---a social media company may be less willing to flag inflammatory language when the speaker is criticizing their own \cite{AllanHateSpeach}.

Social media's endlessly logged interactions have also been a boon to understanding human behavior. Researchers have used various social networks to model bullying \cite{cheng2015antisocial}, urban mobility \cite{noulas2012tale}, and the interplay of friendship and shared interests \cite{yang2011like}. Such studies do not have the benefit of a controlled setting where a single variable can be isolated. However, orders of magnitude more observations in participants' natural habitat offer more fidelity to lived experience \cite{kosinski2015facebook}. Additionally subjects can be sampled from countries not so singularly Western, Educated, Industrialized, Rich, and Democratic---or WEIRD, in the parlance of Henrick et al \cite{henrich_heine_norenzayan_2010}.

In this paper we show how readily different personality and demographic information can be extracted from Facebook statuses. Our reported performance is useful to learn how traits are related to online behavior. For example, sensational interests as measured by the Sensational Interest Questionnaire (SIQ) have been studied for internal reliability \cite{egan1999sensational}, relationship to physical aggression \cite{egan2009sensational}, and role in intrasexual competition \cite{weiss2004sensational}. Yet work connecting SIQ with social media use relies on individually labeling sensational interests in statuses and is only predictive among males \cite{hagger2011social}. Our model performs well for both males and females without hand-labeling statuses. Similarly, other research found no relationship between satisfaction with life (SWL) and status updates \cite{wang2014can}; we show modest test set performance. Finally, although Facebook Likes have been shown to be highly predictive of many personal traits \cite{kosinski2013private}, language models with good performance on this dataset have been limited to predicting personality, age, and gender \cite{schwartz2013personality,farnadi2016computational,sap2014developing}.

The benchmark also helps assess the efficacy of services that explicitly or implicitly rely on inferring these traits. This is valuable to those developing new services as well as to users concerned about privacy. Of particular interest is the role of psychographic models in Cambridge Analytica's (CA) marketing strategy. From leaked internal communications, in 2014 CA amassed a dataset of Facebook profiles and traits almost identical to those in the myPersonality dataset \cite{nyt}. The week after CA's project became public, Facebook's stock plummeted \$75 billion \cite{marketwatch}. One factor in that drop was the belief that Facebook had allowed a third party to create a powerful marketing tool that could manipulate elections \cite{guardianBannon,nyt}.
There are dozens of publications on the myPersonality dataset. However, this is the first to predict SIQ, fair-mindedness, and self-disclosure, which CA discussed in relation to building user models \cite{nyt}.

Besides performance benchmarks, the other major contribution of this paper are the most highly weighted words to predict each trait. The weights also say something about human behavior. The interpretation here is more complex: regression on tens of thousands of features is fraught with over-fitting and colinearity. Despite those problems, in Section \ref{interpret} we argue that the weights can still be treated as a data exploration tool similar to clustering. We provide examples of previously studied relationships that are borne out in the word lists, and believe the lists are a useful tool to develop yet unstudied hypotheses.

Highly weighted features are also an important way to analyze models. We argue in section \ref{CA} that a militarism predictor CA may have built is accurate, but extracts obvious features. Additionally, by inspecting the features in an Atheist vs. Agnostic classifier we find many gendered words. We demonstrate the bias empirically, then fix the classifier to be more fair. This approach is instructive for interrogating more critical models built on social media data.

This paper includes many contributions that could stand alone. We show that the text of Facebook statuses can predict user SWL and SIQ. We expand the prediction of political identity from a single spectrum (liberal/conservative) to twelve distinct ideologies with varying levels of overlap and popularity. On that task, we establish state of the art performance with a model that also provides informative features for every pairwise political comparison. We recreate models CA may have built, and report their performance and the type of information they extracted. We bring character level deep learning to gender prediction. To our knowledge, we also set the standard for predicting IQ, fair-mindedness, self-disclosure, race, and religion from Facebook statuses. Finally, we propose a novel method to make classification less biased.

Given the broad scope of this paper, some contributions are given less space than they would typically merit. Even so, we believe it is important to report results on many traits in a single paper. This demonstrates the power of a simple model and allows task difficulty and extracted features to be compared across traits without concerns about changing experimental setup. 

\vspace{-10pt}
\section{Background}
\vspace{-5pt}
\subsection{myPersonality Dataset}
\noindent From 2008 to 2012, over 7 million Facebook users took the myPersonality quiz produced by the psychologist David Stillwell \cite{kosinski2015facebook}. After answering at least 20 questions, users were scored on the Big Five personality axes: openness, creativity, extraversion, agreeableness, and neuroticism. Over 3 million of those users agreed to give researchers access to their extant Facebook profile and their personality scores. A much smaller subset of users answered additional questionnaires about their interests, Friends' personality, belief in astrology, and other personal information. The research community has added to the dataset by providing race labels for several hundred thousand users; representing the text of statuses in terms of their Linguistic Inquiry and Word Count (LIWC) statistics \cite{pennebaker2001linguistic}; and much more. Labels used in this study are listed in Tables \ref{acc_cat} and \ref{acc_cont}, along with descriptive statistics. To see all available labels, visit myPersonality.org.

myPersonality.org lists 43 publications that use this data. Most work explores the relationship between personality and easily extractable features such as number of Friends or Likes, geographic location, or user-Like pairs. For example, user-Like pairs are shown to be better predictors of a personality than one's spouse \cite{youyou2015computer}. In 2013, Schwartz et al introduced the open vocabulary approach (or bag of words) to personality, gender, and age prediction \cite{schwartz2013personality}. This significantly outperforms closed-vocabulary approaches such as LIWC that rely on domain knowledge to assign each word to one or more of 69 categories. For an excellent overview of related work, we direct readers to that paper's introduction \cite{schwartz2013personality}.

\vspace{-7pt}
\subsection{Language Models}
\vspace{-2pt}
\subsubsection{Bag of Words}
\noindent The majority of our experiments use bag of words (BoW) term frequency-inverse document frequency (tf-idf) preprocessing followed by $\ell_2$ regularized regression. First, the vocabulary is limited to the $k$ most common words in a given training set. Then a matrix of word counts, $N$, is constructed, where $N_{ij}$ refers to how often word $j$ is used by subject $i$. Each row is normalized to sum to one, moved to a log scale, and divided by $d$, the ratio of documents in which each word appears. In more formal notation, each element of the tf-idf matrix is defined by

\vspace{-10pt}
$$W_{ij} = \frac{1 + \log\Big(\frac{N_{ij}}{\sum_{i=1}^{k}N_{ij}}\Big)}{d_j}.$$

\noindent $W$ is then normalized so each row lies on the unit sphere. $W$ can now be used for linear classification or regression with $\ell_2$ regularization on the parameters. This is commonly called Ridge Regression. For binary classification problems, labels are assigned values of $\{-1,1\}$ and a threshold determines predicted label. For categorical data with more than two labels, we train a classifier on each pair of labels. Predicted label is decided by majority vote of the $\frac{c(c-1)}{2}$ classifiers, where $c$ is the number of classes.

\subsubsection{Character-Level Convolutional Neural Network}\label{CNN}
For gender prediction, we also train a 49 layer character level convolutional neural network (char-CNN)  described in \cite{conneau2017very}. Much like successful computer vision architectures \cite{krizhevsky2012imagenet}, each character is embedded in continuous space and combined with neighbors by many layers of convolutional filters. Unlike BoW models, CNNs preserve the temporal dimension, allowing the use of syntactic information. While a great advantage, and theoretically more similar to human cognition, this requires different preprocessing. During training, all inputs must be the same length along the temporal axis despite the wide variation in total length of users' statuses. We chose to split users' concatenated statuses into chunks of no more than 4000 characters, and no less than 1000, as this is enough text for humans to perform gender classification \cite{nguyen2014gender}. Each chunk contains roughly 800 words. Chunks from the same user are assigned entirely to either the training or test set. Unfortunately, preprocessing differences do not allow for a direct comparison between methods. However, enforcing the same preprocessing for both models would necessarily limit one.

\vspace{-5pt}
\subsection{Labels}
Tables \ref{acc_cont} and \ref{acc_cat} provide statistics of the continuous and categorical data respectively. What follows is a brief description of each label and how it was collected.

\def\spc{-10pt}
\vspace*{\spc}
\subsubsection{Gender} is the binary label users supplied when setting up their Facebook account. Offering this information was common before 2008, and mandatory from 2008-2014. In 2014, (after the collection of this dataset) Facebook added 56 more gender options but still uses a binary representation to monetize users \cite{bivens2017gender}.

\vspace*{-10pt}
\subsubsection{Race} labels provided in the dataset are inferred from profile pictures using the Faceplusplus.com algorithm which can identify races termed White, Black, and Asian. A noisy measure of visual phenotype is not the gold standard for the study of race, however, our results indicate it is related to social media use.

\vspace*{-10pt}
\subsubsection{Political identity} is limited to the twelve most common responses: IPA, anarchist, centrist, conservative, democrat, doesn't care, hates politics, independent, liberal, libertarian, republican, and very liberal. These are heterogenous categories from an open-ended question. No work was done to limit labels to political parties (eg. remove ``doesn't care''), disambiguate misspelled or similar responses (eg. combine ``anarchy'' and ``anarchist'' or ``liberal'' and ``very liberal''), or limit responses to one country. To produce the word list for Liberals and Conservatives in Table \ref{words_rel_pol}, we combine ``liberal'', ``very liberal'', and ``democrat’’ as well as ``conservative'', ``very conservative'', and ``republican''. The most likely meaning of IPA is the Independence Party of America, which was in its nascence during this survey. The party is most popular among young people disaffected by the two party system, a sentiment reflected by the users who report IPA.

\vspace*{-10pt}
\subsubsection{Religion} categories were limited to the nine most common responses, and similar labels were combined. Three variants of Catholic---``catholic'',``christian-catholic'', and ``romancatholic''---were merged to form Catholic. Likewise, Christian refers to ``christian'', ``christian-baptist'' and ``christian-evangelical''. The entire list includes: Atheist, Agnostic, Catholic, Christian, Hindu, and None.

\vspace*{-10pt}
\subsubsection{Belief in star sign} is the user's response to ``Horoscopes provide useful information to help guide my decisions?'' Options include: Strongly Agree, Slightly Agree, No Opinion, Slightly Disagree, and Strongly Disagree.

\vspace*{-10pt}
\subsubsection{Personality} is determined on five axes---Openness, Conscientiousness, Extroversion, Agreeableness, and Neurotocism---by a survey. Users answer 20-300 questions which are used to score each personality component on a scale of 1-5. There is a large body of research showing that five factor analysis is explanatory for behavior \cite{digman1990personality}, and its measurement is reproducible \cite{mccrae1987validation}. That work is now adapting to larger datasets collected online \cite{kosinski2015facebook}.

\vspace*{-10pt}
\subsubsection{Sensational Interests} include Militarism, Violent-Occult, Intellectual Recreation, Occult Credulousness, and Wholesome activities. Users can indicate ``Great Dislike'', ``Slight Dislike'', ``No Opinion'', ``Slight Interest'', and ``Great Interest'' for 28 different items including: ``Drugs'', ``Paganism'', ``Philosophy'', ``Survivalism'', and ``Vampires and Wolves''. Interest levels are calculated by summing responses from relevant items. The full calculation can be found in \cite{egan1999sensational}.

\vspace*{-10pt}
\subsubsection{IQ} is determined by 20 questions that conform to Raven's Standard Progressive Matrices. The development and validation of these questions is explained in \cite{IQkosinski} and \cite{kosinski2014measurement}. Because performance on IQ tests has been rising at roughly 0.3 points a year over the past century and IQ is defined as mean 100, the scoring of a test is properly defined over an age cohort \cite{flynn1987massive}. These scores do not take age into account and the mean is 114.

\vspace*{-10pt}
\subsubsection{Satisfaction with life, self-disclosure, and fair-mindedness} are assessed by separate questionnaires. SWL is a measure of global well being somewhat robust to short term mood fluctuations \cite{diener1985satisfaction}.

\section{The Interpretation of Feature Weights}\label{interpret}

A common approach to understand traits in social science is to solve

$$X = UT + \epsilon,$$
where $X$ is observations of subjects, $T$ is the traits of subjects, $U$ is a transition matrix, and $\epsilon$ is model error \cite{khandani2010consumer,egan1999sensational,cooke2004demographic,pecina2013personality,quilty2009personality,tett1991personality,park2015automatic,cesare2017detection,kleinberg2016inherent}. Traits are preferred to be orthogonal to promote compactness without sacrificing modeling power. The Big 5 personality model is both criticized and defended on grounds of trait independence, explanatory power, and measureability, which conforms to the linear model above \cite{john1999big}. Because the traits are defined by language they will not be completely orthogonal. Additionally, observations are not independent. As such, values in $U$ will have dependencies across both rows and columns. Some traits like personality are used to predict other traits or life events \cite{egan1999sensational,tett1991personality}. Learning those relationships can be interpreted as informing our beliefs about column dependencies for $U$ when both traits are part of $T$.

In this paper, $X$ is the tf-idf word matrix, $T$ is defined by our labels, and the model weights are some estimate of $U$ we define as $\hat{U}$. Row dependencies in $\hat{U}$ are based on how words function. For example, `camp' and `camping' perform similar roles in a status. Likewise, the relationship between IQ and agreeableness will be embedded in the columns of $\hat{U}$. However, many of the tasks have little training data and the solution is ill-posed. Regularization encourages generalization, but does not provide any guarantees. Further, sometimes $\epsilon$ dominates the model when observations are not very explanatory or the relationship to a trait is not linear. Given these challenges, what confidence can be placed in the estimate $\hat{U}$?

These problems mirror those faced when clustering data. Clustering does not come with guarantees it will yield sensible answers in diverse scenarios \cite{kleinberg2003impossibility}. However, it is broadly useful when exploring large sets of data \cite{jain1999data,shamir20021,dixon2003classification}. Similarly, $\hat{U}$ can be viewed as a way of ranking features for exploration. A highly ranked observation is not proof it is important. But several highly ranked observations with functional coherence may suggest a hypothesis; particularly when coupled with domain knowledge of row and column dependencies in $U$.

The 55 most highly weighted features for each label are reported in the Appendix. Though the word lists are shown in order of importance, this ranking is not strict. Different regularization, preprocessing, or train/test splits can alter the ordering, especially when there are few examples. Additionally, more common words with lower weights may be used more often in a model's prediction, but may not appear at the top of a list. One may use $\ell_1$ regularization to obtain an arbitrary small number of non-zero weights \cite{meinshausen2009lasso}. This encourages weighting common words and provides more stable rankings. We demonstrate that approach with our IQ model in Section \ref{IQ}.

There are many well-studied phenomena embedded in the $\hat{U}$ produced by our work. For example, Sarah Palin is the only politician indicated in the liberal word list in Table \ref{words_rel_pol}. Likewise, Nancy Pelosi ranks just below Ronald Reagan among conservative words. This accords with literature on the memorability of negative ads \cite{lau2007effects}, importance of outgroup prejudice for social identity \cite{huddy2003group,branscombe1994collective}, and biases women face in politics \cite{schneider2014measuring,dolan2010impact}. We hope the many word lists in the appendix will be useful to researchers in the development of new hypotheses.

$\hat{U}$ is also useful to understand models built on social media data. 
Until recently, the models themselves were not very important. However, machine learning can now be used to estimate sensitive traits such criminal recidivism \cite{kleinberg2016inherent}. 
Given the literalness with which estimates are often interpreted, it is essential to note that model weights are causal for the predicted label. In Section \ref{gender_bias} we use our understanding of the input features to characterize information the model extracts to predict religion. This dataset also includes demographic labels, which show predicted religion labels are more gendered than the ground truth.

We hope the included word lists (a) highlight unstudied relationships about these traits (b) illustrate what kind of information is extracted from social media by machine learning systems.

\section{Results and Discussion}

\subsection{Experimental Setup}
All BoW experiments employ the same preprocessing. Users must have over 500 words in the sum of all their statuses. 80\% of the data is randomly assigned to the training set; the remaining samples constitute the test set. The vocabulary is limited to the 40,000 most common words in each training set. Words must be used by at least 10 users but no more than 60\% of users in the training set. The regularization parameter is tuned via efficient leave one out cross validation \cite{vehtari2015efficient} when $n<10,000$, and $3$-fold cross validation for larger datasets. All BoW models are implemented using the sklearn library \cite{scikit-learn}. Table \ref{acc_cont} reports the number of samples and explained variance (EV) of the predictions on continuous data. Table \ref{acc_cat} reports the number of classes, ratio of samples in the dominant class, homogeneity, and performance on tasks with categorical data.

\begin{table}[!hbt]
\centering
\caption{Prediction Accuracy on Continuous Data}
\label{acc_cont}
\begin{tabular}{lll}
\textbf{Label}                 & \textbf{N} & \textbf{EV} \\ \hline
\textbf{Personality}           &            &             \\
~Openness                       & 84451      & 0.171       \\
~Conscientiousness              & 84451      & 0.120       \\
~Extroversion                   & 84451      & 0.141       \\
~Agreeableness                  & 84451      & 0.090       \\
~Neuroticism                    & 84451      & 0.100       \\
\textbf{Sensational Interests} &            &             \\
~Militarism                     & 4074       & 0.165       \\
~Violent-Occult                 & 4074       & 0.192       \\
~Intellectual Recreation        & 4074       & 0.033       \\
~Occult Credulousness           & 4074       & 0.144       \\
~Wholesome Activities           & 4074       & 0.108       \\
Satisfaction With Life         & 2502       & 0.034       \\
Self Disclosure                & 2006       & 0.092       \\
Fair-Mindedness                & 2006       & 0.064       \\
IQ                             & 1807       & 0.128      
\end{tabular}
\begin{flushleft}
Explained Variance (EV) is 1-$\frac{\mathrm{Var}(y-\hat{y})}{\mathrm{Var}(y)}$, where $\hat{y}$ is the predicted label.
\end{flushleft}
\end{table}

\begin{table}[]
\centering
\caption{Prediction Accuracy on Categorical Data}
\label{acc_cat}
\begin{tabular}{lllllll}
\textbf{Label}      & \textbf{N} & \textbf{Classes} & \textbf{Mode} & \textbf{Homogeneity} & \textbf{F1-score} & \textbf{Acc} \\ \hline
Gender              & 109104     & 2                & 0.598               & 0.519                & 0.92        & 0.903        \\
Race                & 22059      & 3                & 0.682               & 0.52                 & 0.74        & 0.766        \\
Political identity  & 19769      & 12               & 0.213               & 0.133                & 0.33        & 0.337        \\
Religious identity  & 8388       & 5                & 0.488               & 0.318                & 0.54        & 0.541        \\
Belief in Star Sign & 7115       & 5                & 0.331               & 0.245                & 0.32        & 0.334       
\end{tabular}
\begin{flushleft}
Mode is the ratio of the dominant class. Homogeneity is the probability two random samples will be of the same class. The F1-Score is the harmonic mean of precision and recall. For non-binary labels, the precision and recall for each class is weighted by its support.
\end{flushleft}
\end{table}

\begin{table}[!hbt]
\centering
\caption{Gender Prediction}
\label{acc_gender}
\begin{tabular}{ll}
\textbf{Model}    & \textbf{Accuracy} \\ \hline
Human Majority Vote & 0.840             \\
LIWC              & 0.784             \\
Tri-grams         & 0.914             \\
Tri-grams + LIWC  & 0.916             \\
BoW (40k Vocab)   & 0.903             \\
BoW (500k Vocab)  & 0.928             \\
49 layer char-CNN & 0.901 
\end{tabular}
\medskip
\begin{flushleft}
Human baseline is the majority vote (n=210) in gender prediction on Twitter data \cite{nguyen2014gender}. LIWC and Tri-grams are reported in \cite{schwartz2013personality}.
\end{flushleft}
\end{table}
\vspace{-10pt}

\subsection{Performance}\label{accuracies}
\subsubsection{Gender}
Table \ref{acc_gender} compares our gender predictor to several other methods. The BoW model with a vocabulary of 500,000 yields accuracy of 92.8\%, 1.4\% more accurate than the tri-gram model reported by Schwartz et al \cite{schwartz2013personality}. Even though the same dataset is used, the comparison is not direct. The tri-gram model seeks to remove the age information from words, has a larger vocabulary, preserves some temporal relationships in the tri-grams, and draws a different train/test split. Moreover, the preprocessing is more restrictive and only includes users with at least 1000 words. Notwithstanding these discrepancies, which may boost or dampen performance, the results are very similar. When the LIWC representation is added to the tri-grams, there is a slight improvement to 91.6\% accuracy. Preprocessing is even less similar for the char-CNN described in the Section \ref{CNN}. The human baseline of 84.0\% consists of volunteer judgments based on 20-40 user tweets as reported by Nguyen et al \cite{nguyen2014gender}. This is less text than is available to the other models, and from a different social media platform. But, with 210 volunteer guesses per user, it provides a relevant human baseline.

\vspace{-10pt}
\subsubsection{Personality}
After gender, personality is the most studied trait in this paper. Likewise, Schwartz et al achieve the best results to date \cite{schwartz2013personality}. They report the square root of EV to two significant digits: 0.42, 0.35, 0.38, 0.31, 0.31. In that format, we are just 0.01 beneath the state of the art for openness and agreeableness, 0.01 better for neuroticism, and equivalent for the remaining traits. As with gender, we achieve this with a simpler model. 

\vspace{-10pt}
\subsubsection{Political Identity}
Prediction accuracy of 33.7\% is a gain of 11.7\% over the baseline strategy of always predicting the mode, `doesn't care'. As noted in the experiments section, training samples are weighted inversely to their class representation; therefore, ignoring any class will result in an equal loss. This does not provide the highest classification accuracy. However, we believe when some classes are sparsely populated an MSE optimal classifier that is highly biased toward the mode should not be the standard. For reference, equal sample weights and the same training scheme yield classification accuracy of 36.3\% and a weighted f1 score of 31.6\%. Five classes---IPA, hates politics, independent, libertarian, and very liberal---have no representation in the test set predictions. The weighted classifier predicts each class at least once.

According to Preotiuc-Pietro et al., all previous research on predicting political ideology from social media text has used binary labels such as liberal vs conservative or Democrat vs Republican. They broaden the classification task to include seven gradations on the liberal to conservative spectrum \cite{preoctiuc2017beyond}. When predicting ideological tilt from tweets, they achieve a 2.6\% boost over baseline (19.6\%) with BoW follow by logistic regression. Word2Vec feature embeddings \cite{mikolov2013distributed} and multi-target learning with some hand-crafted labels yield an 8.0\% boost. From classification along grades of a single spectrum, we significantly expand the task to twelve diverse identities with varying levels of representation and ideological overlap while maintaining classification accuracy.

In Table \ref{pol_mat} we report the matrix of highest weighted words for separating users in each pairwise class comparison. As with race, belief in star sign, and religion, we plan on making expanded pairwise lists available online. In Table \ref{cm_pol} we report the confusion matrix. Note that many errors are between similar labels, such as liberal and democrat. Ease of training, strong performance, and representation of minority classes make a majority vote system of shallow pairwise classifiers a good approach for this task.

For binary comparison, by pooling \{`very liberal',`liberal',`democrat'\} and \{`very conservative',`conservative',`republican'\} we achieve 76.4\% accuracy; 12.1\% above baseline. Table \ref{words_rel_pol} shows the top 55 liberal and conservative words.

\vspace{-10pt}
\subsubsection{Religion}
Religion seems to be more difficult to glean from statuses than political identity. At 54.1\%, accuracy is a modest 5.3\% above guessing the mode. The most highly weighted pairwise words are on Table \ref{mat_rel}, and Table \ref{cm_rel} shows the confusion matrix. The most highly weighted word to distinguish someone who is agnostic from an atheist is `boyfriend'. This led us to look deeper at that pairwise classifier in Section \ref{gender_bias}. Binary labels were constructed by pooling \{`catholic',  `christian-catholic',  `romancatholic',  `christian',  `christian-baptist'\} and \{`atheist', `agnostic',`none' \}. We achieve 78.0\% accuracy, 5.2\% above baseline. Those words are on table \ref{words_rel_pol}. To our knowledge, there is no other multi class religion predictor to which our results can be compared.

\vspace{-10pt}
\subsubsection{IQ}\label{IQ}
In a genome wide association meta study of 78,308 individuals, 336 single nucleotide polymorphisms were found to explain 2.1-4.8\% of the IQ variance among the test population \cite{sniekers2017genome}. We achieve 12.8\% EV with a model trained on less than 2000 users and their statuses. Using $\ell_1$ regularization to limit the vocabulary to the ten most informative words---final, physics; ayaw, family, friend, heart, lmao, nite, strong, ur---still yields 5.6\% percent EV. The relative accuracy of such a trivial model  that leverages intuitive features is a helpful comparison for any project predicting this important trait. To our knowledge, this is the only work to date that infers IQ from social media. 

The selected features are also informative. Words suggesting intelligence---`final' and `physics'---are parsimonious and singularly academic. Whereas the university experience is sufficient to find users with high IQ, features inversely related to IQ are more focused on disposition. From table \ref{words_big5}, agreeableness is implied by `family' and `heart'; conscientiousness is implied by `family' and `lmao'; and low openness is implied by `ur'. Overall, the list can be characterized as prosocial, or at least concerned with social relationships. Predicting low IQ with prosocial features seems to challenge some previous research.

Gottlieb et al observed that learning disabled children were more likely to engage in solitary play \cite{gottlieb1986sociometric}. Play has also been observed to be more aggressive \cite{bryan1976come}. More directly related to our task, McConaughy and Ritter showed a positive correlation between the IQ of learning disabled boys and social competence scores; and a negative correlation between IQ and behavior problem scores \cite{mcconaughy1986social}. For further review of the subject see \cite{bellanti2000disentangling}.

An MSE optimal classifier seeks to generalize information about samples near the average. This can cause bias when classifying minorities, but is instructive when interpreting features. Features should say something about the majority of our sample, those with IQ near the mean. This explains why antisocial behavior among those with extremely low IQ does not preclude prosocial behavior indicating moderately lower IQ. Reflecting the limitations of this type of study, words like `family', `friend', and `heart' could also be caused by differing norms for social media use or many other factors. Prosocial words predicting lower IQ does however suggest interesting future work.

\vspace{-10pt}
\subsubsection{Sensational Interests}
In this study, SIQ is the easiest continuous variable to predict, even with an order of magnitude less training data than personality. The SIQ asks lists 28 discrete interests like `black magic' and `the armed forces'. Very similar terms can be recovered from statuses: `zombie', `blood', `vampire'; `military', `marines', `training'. Personality tests, on the other hand, ask more abstract questions like `I shirk my duties' for conscientiousness. Many of these duties seem to be extracted in Table \ref{words_big5}: `studying', `busy',`obstacles'. But many more training examples are required for similar performance. 

This is the first work to demonstrate an automatic system for predicting SIQ. Previous research relied on manually counting the number of sensational interests in statuses. The count was only correlated with militarism among men; the relationship was negative for women \cite{hagger2011social}.

\vspace{-10pt}
\subsubsection{Satisfaction With Life}
Previous research cast doubt on the relationship between status updates and SWL \cite{wang2014can}. The number of positive words used on Facebook nationwide in a given day, week, or month, is inversely correlated with the SWL of that time period’s myPersonality participants. The interpretation of that result  is that it ``challenges the assumption that linguistic analysis of internet messages is related to underlying psychological states.'' Here we show that a BoW model accounts for 3.4\% of the variance in SWL scores. Moreover, the most important words the model finds are intuitive. Lower SWL is implied by ``fucking'', ``hate'', ``bored'', ``interview'', ``sick'', ``hospital'', ``insomnia'', ``farmville'', and ``video''. The deleterious effects of joblessness, anger, chronic illness, and isolation are well documented. Words positively associated with SWL---``camping'', ``imagination'', ``epic'', ``cleaned'', ``success''---make similar sense.

Conversational AI on Facebook Messenger is an efficacious and scalable way to administer cognitive behavioral therapy \cite{fitzpatrick2017delivering}. Our results show linguistic analysis can shed light on underlying psychological states. This is important to find users that could benefit from such treatment.


\subsubsection{Belief in Star Sign}
Compared to political identity, BSS has seven fewer classes and a far more homogeneous distribution. Even so, the BSS classifier performs slightly worse than the politics classifier and roughly on par to the baseline of predicting the mode. Unlike our race, gender, politics and sensational interests, we don't wear belief in astrology on our sleeve.




\subsection{Model Selection} 
\noindent BoW models are somewhat unintuitive. Humans use syntactic information when decoding language, which the model discards. Yet, for many tasks they achieve state of the art performance. We compare our BoW to a character-level CNN on gender prediction, our most data rich problem. A character-level CNN is well suited to large amounts of messy, user generated data. Pooling layers in a CNN allow generalization of words like ``gooooooooo'' and ``gooooooo'', while BoW must learn distinct weights. Surprisingly, the CNN does not outperform the simple BoW as shown in Table \ref{acc_gender}.

We found the choice of prediction model is not as important as preprocessing. In initial experiments, Support Vector Machines \cite{suykens1999least} and logistic regression, and $\ell_2$ regularized regression yielded similar performance, depending on choice of $n$-grams and whether Singular Value Decomposition was used \cite{golub1970singular}. We implement ridge regression and classification for simplicity.

Inferring human traits from social media is now being done using deep models \cite{iyyer2014political,preoctiuc2017beyond}. That may be useful in some cases, but for this project the deep model offered no performance boost or intuition to underlying human behavior. Perhaps a continuous bag of words \cite{mikolov2013distributed} and recurrent neural network \cite{felbo2017using} would have done better, but researchers should not consider deep learning essential for this field. Moreover, any performance gains should be weighed against loss of interpretability.

\subsection{Cambridge Analytica} \label{CA}
With current technology, Facebook statuses are a better predictor of someone's IQ than the totality of their genetic material \cite{sniekers2017genome}. When a marketing firm adds such a tool to their arsenal it is natural to be suspicious. Indeed, The Guardian article that broke the CA story was headlined ```I made Steve Bannon's psychological warfare tool': meet the data war whistleblower'' \cite{guardianBannon}. (Steve Bannon is the former chief executive of the Trump presidential campaign.) However, closer inspection of psychographic models casts doubt on their ability to add value to an advertising campaign, even when the predictions are accurate. In this paper we show that militarism is one of the most easily inferred traits. At 16.5\% explained variance, it is more predictable than any of the big 5 personality traits except openness, even with just 5\% of the training data. SIQ is also a much stronger predictor of aggressive behavior than the Big 5 \cite{egan2009sensational}. If this trait was actionable for the Trump campaign, it is interesting that the two most highly weighted features are `xbox' and `man'. Gaming interest and gender are already available via Facebook's advertising platform; reaching that demographic does not require an independent model. Additionally, Steve Bannon's belief in the political power of gamers predates CA's psychographic model by a decade \cite{wiredBannon}.

Readers are encouraged to view the word lists in the Appendix through the lens of task accuracy on Tables \ref{acc_cont} and \ref{acc_cat}. They may come to the same conclusion as the Trump campaign who, according to CBS News, ``never used the psychographic data at the heart of a whistleblower who once worked to help acquire the data's reporting -- principally because it was relatively new and of suspect quality and value.'' \cite{CBSbig5}. Performance results and extracted features allow for more informed discussion; particularly for SIQ, fair-mindedness and self-disclosure on which we report the first accurate prediction model.

There are limitations to this analysis. Our models only use statuses; Likes and network statistics could increase accuracy. Further, other psychographic traits beyond militarism may be politically useful but have no obvious demographic stand-in. Finally, we don't have access to CA's exact dataset and instead built our models on the myPersonality dataset.

\begin{table}[!hbt]
\caption{Agnostic vs Atheist Confusion Matrix}
\label{cm_biased}
\begin{minipage}{.5\textwidth}
\centering
\begin{tabular}{lllll}
                                                   &                   & \multicolumn{2}{c}{\textbf{Predicted (Men)}} &                \\
                                                   &                   & \textbf{Agnostic}  & \textbf{Atheist}  & \textbf{Total} \\
\multicolumn{1}{r}{\multirow{2}{*}{\textbf{True}}} & \textbf{Agnostic} & 36                 & 33                & 69             \\
\multicolumn{1}{r}{}                               & \textbf{Atheist}  & 28                 & 58                & 86             \\
                                                   & \textbf{Total}    & 64                 & 91                &               
\end{tabular}
\end{minipage}%
\begin{minipage}{.5\textwidth}
\centering
\begin{tabular}{lllll}
                                                   &                   & \multicolumn{2}{c}{\textbf{Predicted (Women)}} &                \\
                                                   &                   & \textbf{Agnostic}  & \textbf{Atheist}  & \textbf{Total} \\
\multicolumn{1}{r}{\multirow{2}{*}{}} & \textbf{} & 86              & 21             & 107            \\
\multicolumn{1}{r}{}                 & \textbf{}  & 34              & 16             & 50             \\
                                   & \textbf{}    & 120             & 37               &               
\end{tabular}
\end{minipage}
\end{table}

\begin{table}[!hbt]
\caption{Fair Agnostic vs Atheist Confusion Matrix}
\label{cm_fair}
\begin{minipage}{.5\textwidth}
\centering
\begin{tabular}{lllll}
                                                   &                   & \multicolumn{2}{c}{\textbf{Predicted (Men)}} &                \\
                                                   &                   & \textbf{Agnostic}  & \textbf{Atheist}  & \textbf{Total} \\
\multicolumn{1}{r}{\multirow{2}{*}{\textbf{True}}} & \textbf{Agnostic} & 40                 & 29                & 69             \\
\multicolumn{1}{r}{}                               & \textbf{Atheist}  & 31                 & 55                & 86             \\
                                                   & \textbf{Total}    & 71                 & 84                &               
\end{tabular}
\end{minipage}%
\begin{minipage}{.5\textwidth}
\centering
\begin{tabular}{lllll}
                                                   &                   & \multicolumn{2}{c}{\textbf{Predicted (Women)}} &                \\
                                                   &                   & \textbf{Agnostic}  & \textbf{Atheist}  & \textbf{Total} \\
\multicolumn{1}{r}{\multirow{2}{*}{}} & \textbf{} & 85              & 22             & 107            \\
\multicolumn{1}{r}{}                               & \textbf{}  & 31              & 19             & 50             \\
                                                   & \textbf{}    & 116                & 41                &               
\end{tabular}
\end{minipage}
\end{table}

\subsection{Gender Bias in Atheist vs Agnostic Classifier}\label{gender_bias}

Highly weighted atheist words include ``fucking'', ``bloody'', ``maths'', ``degrees'', ``disease'', ``wifey'', and ``religion''. Meanwhile, ``beautiful'', ``santa'', ``friggin'', ``thank'', ``hubby'', ``miles'', and ``paperwork'' imply the user is agnostic. This paints a picture of academic, male, disagreeable and British atheists. Agnostic words are more positive, female, and related to mundane preparation.
A more complete list is shown in Table \ref{words_rel_pol}. What follows is an empirical analysis of our estimator`s gender bias, a discussion of fairness, and results debiasing the model.

In this dataset, atheists and agnostics are 33.5\% and 50.3\% female respectively. This is a stronger female preference for agnosticism than random surveys across the United States which report 32\% and 38\%, respectively \cite{pew}. Table \ref{cm_biased} shows the confusion matrices for men and women. The ratio of predicted to true agnostics is 0.945 for men and 1.35 for women. Similarly, the ratio of false atheist to false agnostic predictions is 90.8\% larger for men than women. The classification of women, the minority in this dataset, is highly distorted.

Models built to generalize information often amplify biases in training data. Cooking videos elicit female pronouns in machine-generated captions 68\% more than male pronouns, even though the training shows only 33\% more women cooking \cite{zhao2017men}. Word embeddings used in machine translation \cite{zou2013bilingual}, information retrieval \cite{clinchant2013aggregating}, and student grade prediction \cite{luo2015predicting} produce analogies such as ``man is to computer programmer as woman is to homemaker''\cite{bolukbasi2016man}. 

There are many notions of fairness defined over an individual \cite{dwork2012fairness,joseph2016rawlsian,kusner2017counterfactual}, population \cite{zafar2017fairness,hardt2016equality}, or information available to the model \cite{grgic2016case}. Building a fair estimator often requires domain knowledge to define a similarity metric \cite{dwork2012fairness}, make corpus-level constraints \cite{zhao2017men}, or construct a causal model that separates protected information from other latent variables \cite{kusner2017counterfactual}. In this paper, we will use the notion of Disparate Mistreatment to measure fairness \cite{zafar2017fairness}. That is, if protected classes experience disparate rates of false positive, false negative or overall misclassification, the estimator is unfair.

To mitigate Disparate Mistreatment we explicitly encode gender---\{$-1$,0,$1$\} for \{male, unknown, female\}---in the feature vector during train time. At test time the gender of all samples is encoded as unknown. The intuition is that latent variables are amplified when they are easy to extract and correlated with the target. As demonstrated by the accuracy of our race and gender predictors, that is often the case for protected information. There often exist more informative, if more subtle, traits than the protected features. For example, atheists and agnostics report a yawning gap in those that don’t believe in God, at 92\% and 41\% \cite{pew}. Additionally, religiosity is shown to be correlated with both Agreeableness and Conscientiousness \cite{saroglou2010religiousness}. But gender is much easier to extract then belief in God or personality. By explicitly giving the model gender information, we hope that the model will do more to extract those other features.

This approach produces much less Disparate Mistreatment of men and women. The ratio of predicted to true agnostics moves closer to parity at 1.02 for men and 1.22 for women. Additionally, the ratio of false atheist to false agnostic predictions is now only 31.8\% larger for men, compared to 90.8\% without intervention. The most highly weighted agnostic words for the new fair classifier are also less gendered; ``hair'', ``wifey'', and ``boyfriend'' are no longer in the top 55, as reported in Table \ref{words_rel_pol}. We also saw no decay in classification rate.

The gender bias of the atheism classifier is clear by simply inspecting its most heavily weighted features. More opaque models should be subjected to more rigorous inspection for bias.

\section{Conclusion and Future Work}

We match or set the state of the art for the 20 traits in this paper. Additionally, we provide the top words for many pairwise classification problems, and top 55 words for regression or binary classification problems. We hope researchers from many fields find the benchmarks and word lists useful. Our analysis of psychographic models in marketing as well as gender bias in a religion classifier are examples of how these performance measures and extracted features can be used together.

In future work we hope to explore what types of unfairness can be solved by our approach in Section \ref{gender_bias}. Further, models built on traits with few examples are well suited to be augmented by transfer learning. This is especially pressing for detecting states like low satisfaction with life, which can be somewhat ameliorated at low cost.

\newpage

\bibliographystyle{ieeetr}
\bibliography{big5.bib}

\newpage
\appendix


\begin{table}[]
\centering
\caption{Pairwise Politics Words}
\label{pol_mat}
\begin{adjustbox}{angle=270}
\begin{tabular}{lllllllllllll}
                                             & \textbf{IPA} & \textbf{anarchist} & \textbf{centrist} & \textbf{conserv.} & \textbf{dem.} & \textbf{doesn't care} & \textbf{hates pol.} & \textbf{indep.} & \textbf{lib.} & \textbf{liber.} & \textbf{repub.} & \textbf{v. lib.} \\ \cline{2-13} 
\multicolumn{1}{l|}{\textbf{IPA}}            &              & fuck           & wishes            & wishes           & smh           & yay                & rain             & congrats        & wishes        & money           & church          & damn             \\
\multicolumn{1}{l|}{\textbf{anarchist}}      & excited      &                & wishes            & driving          & excited       & lol                & dont             & driving         & excited       & ready           & ready           & excited          \\
\multicolumn{1}{l|}{\textbf{centrist}}       & xd           & fuck           &                   & lord             & today         & tattoo             & shit             & surgery         & shit          & government      & school          & damn             \\
\multicolumn{1}{l|}{\textbf{conservative}}   & xd           & fuck           & damn              &                  & fb            & anymore            & shit             & damn            & damn          & art             & school          & damn             \\
\multicolumn{1}{l|}{\textbf{democrat}}       & xd           & fuck           & wishes            & tonight          &               & stupid             & fuck             & died            & wishes        & government      & church          & wishes           \\
\multicolumn{1}{l|}{\textbf{doesn't care}}   & packers      & fuck           & wishes            & lord             & smh           &                    & shit             & definitely      & wishes        & government      & church          & damn             \\
\end{tabular}
\end{adjustbox}
\end{table}

\begin{table}[]
\centering
\caption{Politics Confusion Matrix}
\label{cm_pol}
\begin{adjustbox}{angle=270}
\begin{tabular}{llllllllllllll}
                                             & \multicolumn{12}{c}{\textbf{Predicted Label}}                                                                                                                                                                                &                \\
                                             & \textbf{IPA} & \textbf{anar.} & \textbf{centrist} & \textbf{conserv.} & \textbf{dem.} & \textbf{doesn't care} & \textbf{hates pol.} & \textbf{indep.} & \textbf{lib.} & \textbf{liber.} & \textbf{repub.} & \textbf{v. lib.} & \textbf{Total} \\ \cline{2-14} 
\multicolumn{1}{l|}{\textbf{IPA}}            & \textbf{0}   & 2              & 3                 & 3                 & 11            & 18                    & 2                   & 1               & 3             & 1               & 16              & 1                & 61             \\
\multicolumn{1}{l|}{\textbf{anarchist}}      & 0            & \textbf{24}    & 4                 & 3                 & 5             & 21                    & 1                   & 3               & 15            & 5               & 4               & 3                & 88             \\
\multicolumn{1}{l|}{\textbf{centrist}}       & 2            & 9              & \textbf{74}       & 40                & 52            & 66                    & 3                   & 6               & 95            & 7               & 43              & 4                & 401            \\
\multicolumn{1}{l|}{\textbf{conservative}}   & 2            & 5              & 29                & \textbf{113}      & 26            & 31                    & 0                   & 7               & 53            & 5               & 62              & 0                & 333            \\
\end{tabular}
\end{adjustbox}
\end{table}

\begin{table}[]
\centering
\caption{Pairwise Religion Words}
\label{mat_rel}
\begin{tabular}{llllll}
                                        & \textbf{athiest} & \textbf{agnostic} & \textbf{catholic} & \textbf{christian} & \textbf{none} \\ \cline{2-6} 
\multicolumn{1}{l|}{\textbf{athiest}}   &                  & boyfriend         & thank             & church             & lol           \\
\multicolumn{1}{l|}{\textbf{agnostic}}  & fucking          &                   & prayers           & church             & lol           \\
\multicolumn{1}{l|}{\textbf{catholic}}  & fucking          & fucking           &                   & lol                & lol           \\
\multicolumn{1}{l|}{\textbf{christian}} & fucking          & fucking           & mass              &                    & xmas          \\
\multicolumn{1}{l|}{\textbf{none}}      & fucking          & apartment         & god               & church             &              
\end{tabular}
\begin{flushleft}
The most highly weighted word from each pairwise classifier. Word implies top label.
\end{flushleft}
\end{table}

\begin{table}[]
\centering
\caption{Religion Confusion Matrix}
\label{cm_rel}
\begin{tabular}{lllllll}
                                        & \multicolumn{5}{c}{\textbf{Predicted Label}}                                                       &                \\
                                        & \textbf{Atheist} & \textbf{Agnostic} & \textbf{Catholic} & \textbf{Christian} & \textbf{None} & \textbf{Total} \\ \cline{2-7} 
\multicolumn{1}{l|}{\textbf{Atheist}}    & \textbf{68}              & 29                & 17                & 16                 & 21            & 151            \\
\multicolumn{1}{l|}{\textbf{Agnostic}}  & 54              & \textbf{69}                & 27                & 55                 & 11            & 216            \\
\multicolumn{1}{l|}{\textbf{Catholic}}  & 27              & 37                & \textbf{172}               & 130                & 9             & 375            \\
\multicolumn{1}{l|}{\textbf{Christian}} & 35              & 48                & 126               & \textbf{560}                & 26            & 795            \\
\multicolumn{1}{l|}{\textbf{None}}      & 22              & 11                & 19                & 50                 & \textbf{39}            & 141            \\
\multicolumn{1}{l|}{\textbf{Total}}     & 206             & 194               & 361               & 811                & 106           & 1678          
\end{tabular}
\vspace{80pt}
\begin{flushleft}
In the remaining tables the top 55 words are listed in order for each trait.
\end{flushleft}
\end{table}

\vspace{-30pt}
\begin{table}[!htb]
\centering
\vspace{-10pt}
\caption{Personality Words}
\vspace{-5pt}
\label{words_big5}
\begin{tabular}{llllll}
\multicolumn{2}{c}{\textbf{Openness}} & \multicolumn{2}{c}{\textbf{Conscientious}} & \multicolumn{2}{c}{\textbf{Extroversion}} \\
\textbf{-}          & \textbf{+}        & \textbf{-}             & \textbf{+}          & \textbf{-}          & \textbf{+}            \\ \hline
bored               & art               & lost                   & gym                 & internet            & party                 \\
boring              & poetry            & fucking                & ready               & quiet               & guys                  \\
husband             & beautiful         & xd                     & weekend             & bored               & amazing               \\
attitude            & universe          & phone                  & excited             & listening           & audition              \\
shopping            & peace             & im                     & success             & apparently          & baby                  \\
dinner              & poem              & bored                  & finished            & computer            & haha                  \\
\end{tabular}
\end{table}

\begin{table}[]
\centering
\vspace{-10pt}
\caption{Personality Words Continued}
\vspace{-5pt}
\label{words_big5_cont}
\begin{tabular}{llllll}
\multicolumn{2}{c}{\textbf{Agreeable}} & \multicolumn{2}{c}{\textbf{Neurotic}} & \multicolumn{2}{l}{\textbf{Satisfaction With Life}} \\
\textbf{-}       & \textbf{+}          & \textbf{-}         & \textbf{+}       & \textbf{-}              & \textbf{+}                \\ \hline
fucking          & wonderful           & loving             & sick             & bored                   & family                    \\
stupid           & amazing             & girlfriend         & nervous          & fuck                    & loving                    \\
kill             & awesome             & wife               & stressed         & fucking                 & hope                      \\
shopping         & haha                & awesome            & depression       & hates                   & thankful                  \\
shit             & smile               & parties            & depressed        & bday                    & india                     \\
burn             & happiness           & party              & anymore          & apparently              & wonderful                 \\
\end{tabular}
\end{table}

\begin{table}[]
\centering
\vspace{-10pt}
\caption{Sensational Interest Words}
\vspace{-5pt}
\label{words_siq}
\begin{tabular}{llllll}
\multicolumn{2}{c}{\textbf{Militaristic}} & \multicolumn{2}{c}{\textbf{Violent-Occult}} & \multicolumn{2}{c}{\textbf{Intellectual Recreation}} \\
\textbf{-}          & \textbf{+}          & \textbf{-}           & \textbf{+}           & \textbf{-}                & \textbf{+}               \\ \hline
sleeping            & man                 & lord                 & hell                 & im                        & life                     \\
ugh                 & xbox                & pray                 & zombie               & course                    & jon                      \\
sad                 & gets                & cousins              & damn                 & boring                    & beautiful                \\
excited             & gotta               & church               & fuck                 & painful                   & dancing                  \\
lovely              & good                & michael              & bitch                & decision                  & yoga                     \\
oh                  & training            & allah                & ass                  & hurts                     & thankful                 \\
\end{tabular}
\end{table}

\begin{table}[]
\centering
\vspace{-10pt}
\caption{Sensational Interest Words Continued}
\vspace{-5pt}
\label{words_siq_cont}
\begin{tabular}{llllll}
\multicolumn{2}{c}{\textbf{Occult Credulousness}} & \multicolumn{2}{c}{\textbf{Wholesome Activities}} & \multicolumn{2}{c}{\textbf{Belief in Star Sign}} \\
\textbf{-}               & \textbf{+}             & \textbf{-}              & \textbf{+}              & \textbf{No}            & \textbf{Yes}            \\ \hline
church                   & zombie                 & coke                    & woot                    & minutes                & omg                     \\
praise                   & ass                    & michigan                & camping                 & didn                   & im                      \\
jesus                    & bitch                  & stupid                  & fish                    & church                 & ready                   \\
lord                     & halloween              & pathetic                & life                    & praise                 & friend                  \\
bible                    & animal                 & ops                     & yesterday               & jesus                  & mind                    \\
christ                   & sign                   & husband                 & beautiful               & probably               & ass                     \\
\end{tabular}
\end{table}

\begin{table}[]
\centering
\vspace{-10pt}
\caption{Psychographic Words}
\vspace{-5pt}
\label{words_psych}
\begin{tabular}{llllll}
\multicolumn{2}{c}{\textbf{Self-Disclosure}} & \multicolumn{2}{c}{\textbf{Fair-Mindedness}} & \multicolumn{2}{c}{\textbf{IQ}} \\
\textbf{-}           & \textbf{+}            & \textbf{-}            & \textbf{+}           & \textbf{-}      & \textbf{+}    \\ \hline
bored                & family                & bored                 & excited              & nite            & exam          \\
fuck                 & loving                & wat                   & business             & ur              & hours         \\
fucking              & hope                  & soon                  & says                 & lmao            & sigh          \\
hates                & thankful              & dad                   & apartment            & alot            & camping       \\
bday                 & india                 & xd                    & great                & family          & finish        \\
apparently           & wonderful             & stage                 & delicious            & omg             & paper         \\
\end{tabular}
\end{table}

\begin{table}[]
\centering
\vspace{-10pt}
\caption{Religion and Politics Words}
\vspace{-5pt}
\label{words_rel_pol}
\begin{tabular}{llllllll}
\multicolumn{2}{l}{\textbf{Agnostic vs Atheist}} & \multicolumn{2}{l}{\textbf{A. vs A. (Fair)}} & \multicolumn{2}{l}{\textbf{Religious vs Not}} & \multicolumn{2}{l}{\textbf{Conservative vs Liberal}} \\ \hline
extra                        & physics           & miles                    & fucking              & church                & fucking               & church                   & damn                      \\
miles                        & fucking           & working                  & physics              & pray                  & fuck                  & truck                    & happy                     \\
turn                         & snowing           & extra                    & wat                  & prayers               & xmas                  & government               & fb                        \\
\textbf{hair}                & shit              & awhile                   & fuck                 & god                   & damn                  & america                  & smh                       \\
packing                      & wat               & packing                  & bloody               & easter                & shit                  & pray                     & marriage                  \\
awhile                       & write             & turn                     & shit                 & lord                  & bloody                & haha                     & xmas                      \\
insane                       & bloody            & super                    & write                & blessed               & hell                  & prayers                  & chicago                   \\
\end{tabular}
\end{table}

\begin{table}[]
\centering
\vspace{-10pt}
\caption{Race Words}
\vspace{-5pt}
\label{words_race}
\begin{tabular}{llllll}
\multicolumn{2}{c}{\textbf{White vs. Black}} & \multicolumn{2}{c}{\textbf{White vs. Asian}} & \multicolumn{2}{c}{\textbf{Black vs. Asian}} \\ \hline
tonight                 & smh                & tonight               & asian                & smh                  & korea                 \\
dad                     & fb                 & blonde                & tt                   & fb                   & sa                    \\
stupid                  & lord               & town                  & tmr                  & lord                 & na                    \\
exited                  & fam                & fuckin                & korea                & wit                  & asian                 \\
thinks                  & nigeria            & ass                   & chinese              & aint                 & gay                   \\
ends                    & yall               & college               & ng                   & da                   & chinese               \\
journey                 & black              & gas                   & na                   & yall                 & internet              \\
\end{tabular}
\end{table}

\end{document}